\documentclass[prl,twocolumn,showpacs,preprintnumbers,amsmath,amssymb]{revtex4}

\usepackage{graphicx}
\usepackage{dcolumn}
\usepackage{bm}
\usepackage{psfrag}
\usepackage{epsfig}
\usepackage{amsmath}
\usepackage{amssymb}
\usepackage{color}
\usepackage[FIGTOPCAP,raggedright,nooneline]{subfigure}

\newcommand{\bra}[1]{\langle #1 |}
\newcommand{\ket}[1]{| #1 \rangle}

\newcommand{\nn}{\nonumber}

\newcommand{\id}{{\sf 1 \hspace{-0.3ex} \rule{0.1ex}{1.52ex}\rule[-.01ex]{0.3ex}{0.1ex}}}
\newcommand{\ignore}[1]{}

\newcommand{\I}{\ensuremath{\mathrm{i}}}
\begin{document}

\title{A ``Single-Photon'' Transistor in Circuit Quantum Electrodynamics}

\author{Lukas Neumeier}
\email{lukas-neumeier@gmx.de}
\author{Martin Leib}
\author{Michael J. Hartmann}
\email{mh@tum.de}
\affiliation{Technische Universit{\"a}t M{\"u}nchen, Physik Department,
James Franck Str., 85748 Garching, Germany}

\date{\today}

\begin{abstract}
We introduce a circuit quantum electrodynamical setup for a ``single-photon'' transistor.
In our approach photons propagate in two open transmission lines that are coupled via two interacting transmon qubits. The interaction is such that no photons are exchanged between the two transmission lines but a single photon in one line can completely block respectively enable the propagation of photons in the other line.
High on-off ratios can be achieved for feasible experimental parameters. Our approach is inherently scalable as all photon pulses can have the same pulse shape and carrier frequency such that output signals of one transistor can be input signals for a consecutive transistor.
\end{abstract}

\maketitle

%
%
Photons are the most suitable carrier for transmitting information over long distances as they are largely immune to environmental perturbations, and can propagate with very low loss and long-lived coherence in a wide range of media \cite{OBrien09}. The use of photons in information processing however still suffers from the inability to realize controlled, strong interactions between individual photons. To make photons a more versatile information carrier, it is therefore of great importance to conceive means of making photonic signals interact with each other \cite{Kimble08}.
In vacuum, direct photon-photon interactions are absent. Nonetheless, optical signals can influence each other in nonlinear media.
Yet, the quantum regime with interactions between individual photons only becomes accessible for devices where optical nonlinearities exceed incoherent and dissipative processes. Suitable devices therefore require a strong coupling of the photons to the material that mediates the effective photon-photon interactions. Since the coupling of light to matter can be enhanced if light fields are confined to small volumes in space, cavities and one-dimensional waveguides are prime candidates for such devices.   

\begin{figure}
\centering
\includegraphics[width=\columnwidth]{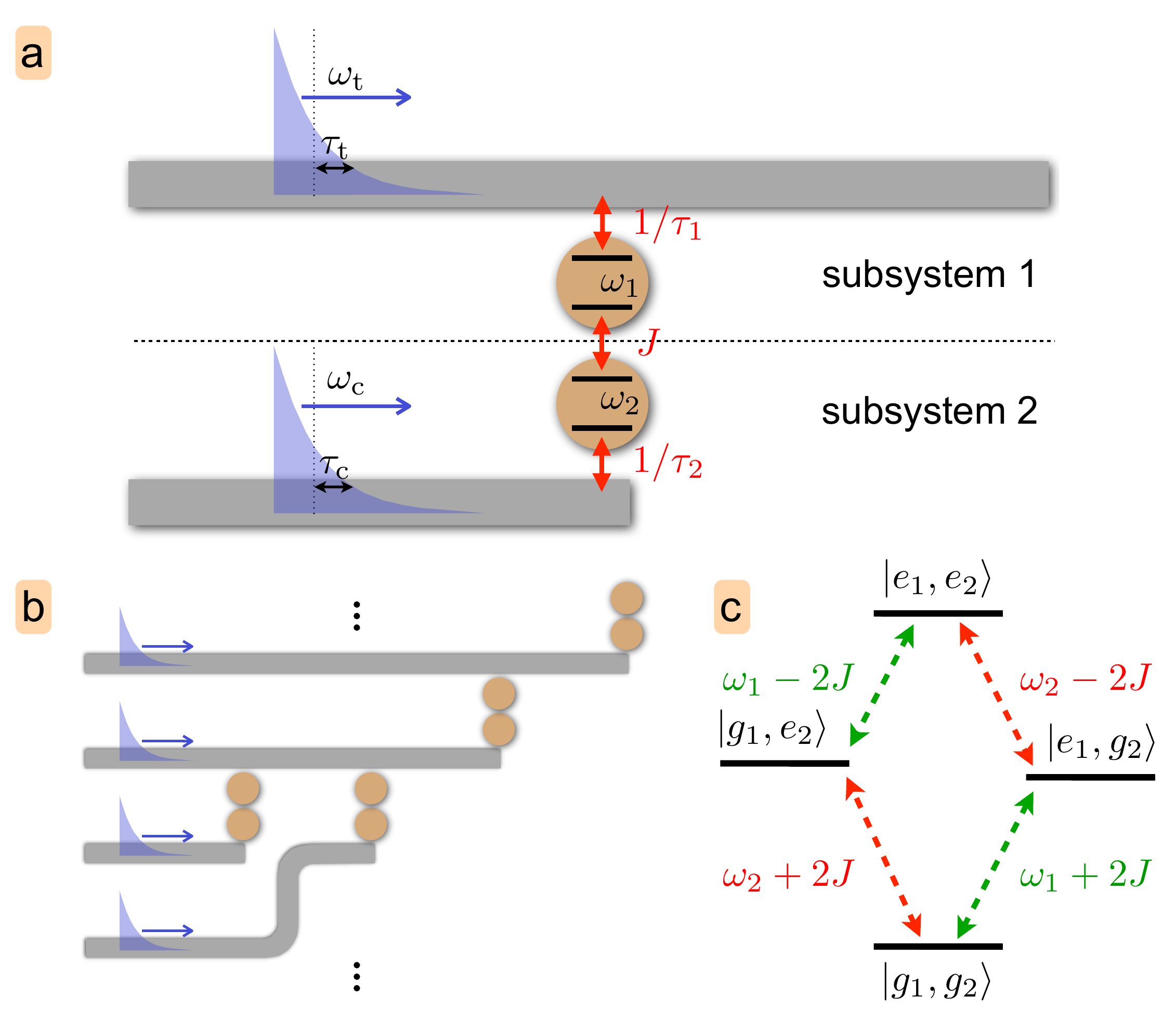}
\caption{Sketch of the considered setup with all pulses sketched as inverting pulses. {\bf a}) Two photons traveling in separate waveguides scatter off each other at a scattering center formed by two interacting qubits. {\bf b}) Multiple devices can be concatenated to form a network. {\bf c}) Level scheme of the scattering center Hamiltonian $H_{sys}$, c.f. equation (\ref{HIsing}).}
\label{pnse}
\end{figure}

Here we introduce a scheme for a ``single-photon'' transistor, a device that can be considered to form a cornerstone of quantum optical information processing. In our approach individual photons propagate in two one-dimensional waveguides of low transverse dimension and scatter off each other at a localized scattering center formed by two two-level systems (qubits) that each couple to one of the waveguides, see figure \ref{pnse}a. The qubits interact in such a way that no excitations can be exchanged between them and thus ensure that each photon remains in its initial waveguide after the scattering event. Nonetheless, as we show below, the presence of a single photon in one waveguide can completely block or enable the propagation of a photon in the other waveguide.
Importantly, our approach works for propagating light signals that all have the same carrier frequency and pulse shape, which makes it inherently scalable as the output signals of one transistor can enter as input signals into a consecutive transistor, c.f. figure \ref{pnse}c for an illustration. Such scalability is questionable in previous proposals which are based on different technological platforms \cite{Chang07,Hong08}.
Moreover the device we propose is a passive element that does not require any temporal tuning of the qubits. This implies that the arrival time of the photons at the scattering center can be completely unknown. Differences between the arrival times of the individual photons do of course matter but the device becomes increasingly insensitive to timing mismatches as qubit dissipation decreases.

A technology that is ideally suited for realizing the device we envision is provided by itinerant microwave photons in superconducting circuits \cite{Wallraff04,Schoelkopf08}.
Here, coherent scattering at a superconducting qubit \cite{Astafiev10,Hoi11,Hoi12} and entanglement with a qubit \cite{Eichler12} have been demonstrated for individual photons that propagate in open transmission lines. Moreover precise shaping of single photon pulses has been shown \cite{Yin12} very recently.
An implementation of our approach in circuit quantum electrodynamics thus requires two superconducting qubits that are coupled to open transmission lines.
We show that the desired qubit-qubit interaction can be realized with two transmon qubits \cite{Koch07} that are coupled via a SQUID which can be tuned to ensure that no excitations are exchanged between both transmons. Importantly, this coupling is not dispersive \cite{Kirchmair12} and thus strong as both transmons can have the same transition frequency. These rather unique possibilities for qubit-qubit interactions offered by superconducting circuits are very suitable for our aims. Moreover, their robustness with respect to dephasing noise make transmons ideal qubits for our device. Yet, alternatively one could also use two flux qubits that are coupled via an induction loop \cite{Orlando00,Mooij99}.

To demonstrate the capabilities of the ``single-photon'' transistor we propose,
we calculate the photon reflection and transmission probabilities for both transmission lines that depend on the incoming photon pulses, under realistic experimental conditions, i.e. taking into account all dissipative processes in our setup.

\paragraph{Setup} \label{sec_master_ham}
We consider two interacting qubits that each couple to a one-dimensional waveguide in which the photons propagate.
Here we focus on a setup for which we can refer to a control and a target photon, where the presence of the control photon influences the target photon's direction of propagation, while the control photon's direction of propagation always changes. A sketch of this setup is shown in figure \ref{pnse}a. 
The control photon propagates in the waveguide of subsystem 2, c.f. figure \ref{pnse}a, which has a closed end right where it couples to qubit 2. This arrangement enhances the absorption of photons by qubit 2 and hence its inversion as compared to an open waveguide end. The target photon in turn propagates in the waveguide of subsystem 1. 
The qubit-qubit interaction is such that no excitations are exchanged between the two qubits which implies that photons can not tunnel between the waveguides. Nonetheless one control photon in waveguide 2 can completely block or enable the propagation of a target photon in waveguide 1.
The Hamiltonian of the two coupled qubits reads,
\begin{equation}\label{HIsing}
H_{sys} = \frac{\omega_1}{2} \sigma_{1}^{z} + \frac{\omega_2}{2} \sigma_{2}^{z} - J \sigma_{1}^{z} \sigma_{2}^{z} \, ,
\end{equation}
where the $\sigma_{i}^{z}$ are Pauli-operators, $\omega_1$ and $\omega_2$ the transition frequencies of the two qubits and $J$ the strength of their mutual interaction. This Hamiltonian can be implemented with two transmon qubits that are coupled via a SQUID, see figure \ref{transmoncircuit} and supplemental material \cite{supplement}, or with two inductively coupled flux qubits \cite{Orlando00,Mooij99}.

Both transmission lines have a continuous spectrum of photonic modes and can be described by the Hamiltonian \cite{Fan},
$H_T = \int^{\infty}_{-\infty} dp \, p \, (r_p^{\dagger} r_p - l_p^{\dagger} l_p) + \int^{\infty}_{-\infty} dp \, p \, b_p^{\dagger} b_p$,
where $r_{p}^\dag$ ($l_{p}^\dag$) creates a photon in subsystem 1 which travels to the right (left) and $b_{p}^\dag$ creates a photon in subsystem 2. $p = v_{g} \eta$, where $v_{g}$ is the group velocity and the wave vector $\eta$ is negative (positive) for left (right) going modes. A semi-infinite transmission line as in subsystem 2
can be described by only one continuum of modes since its incoming and outgoing modes can be mapped to an infinite waveguide where photons only propagate in one direction \cite{spiegelgleicheinemode}.
The dispersion relation of a transmission line is linear and the frequency integration can be extended to $\pm \infty$ since we only consider pulses with a frequency width that is much smaller than their carrier frequency. For these narrow linewidth pulses we thus take the photon-qubit coupling to be independent of the photon frequency,
\begin{equation}\label{HIisikng}
H_I  =  \int^{\infty}_{-\infty} dp \left[\frac{\sigma_{1}^{+}(r_p + l_p)}{\sqrt{2 \pi \tau_1}} + \frac{\sigma_{2}^{+} b_p}{\sqrt{\pi \tau_2}} + \text{H.c.}\right].
\end{equation}
Here $\tau_1$ and $\tau_2$ are the lifetimes of the two level systems associated to their coupling to the transmission lines.
Figure \ref{pnse}c shows the level scheme of the two qubits described by $H_{sys}$ and the transitions induced by the photons.
The total Hamiltonian that includes the transmission lines, the qubits and their couplings thus reads, 
\begin{equation}\label{Ham}
H = H_T + H_{sys} + H_I .
\end{equation}
In a realistic system, the qubits will be subject to dissipation. We thus assume relaxation of excited qubit levels at a rate $\gamma_{\text{r}}$ and pure qubit dephasing at a rate $\gamma_{\varphi}$ to derive quantum Langevin equations \cite{GardinerZoller} for the photon and qubit operators that describe the unitary dynamics generated by $H$ and the dissipative processes associated to $\gamma_{\text{r}}$ and $\gamma_{\varphi}$. The explicit forms of these equations are given in the supplemental material \cite{supplement}. 
To investigate the dynamics of ``single-photon'' pulses in this setup, we combine quantum scattering theory \cite{taylor} with the input-output formalism \cite{Gardiner} of quantum optics as in \cite{Fan}, where the source terms for the input-output relations are provided by the solutions of the mentioned Langevin equations, see supplemental material \cite{supplement} for details.

\paragraph{Photon-Photon Interaction}

To see the effect of the photon-photon interaction most clearly, we first consider the situation in which only one target photon but no control photon is present.
An incoming target photon that travels to the right is described by $\ket{\Psi_{_{\text{t}}}} = \int dk \, \alpha_{_{\text{t}}}(k)\,  r^\dag(k) \ket{0}$, where $k$ labels the frequency components. We assume for the target photon a pulse with a Lorentzian frequency distribution, 
$\alpha_{_{\text{t}}}(k) = \{ \sqrt{\pi \tau_{\text{t}}} \, [\I (\omega_{\text{t}}-k)+\tau_{\text{t}}^{-1}] \}^{-1}$. Here $\tau_{\text{t}}$ is the temporal width of the pulse and $\omega_{\text{t}}$ its carrier frequency.
A pulse of this form would for example describe a photon that was spontaneously emitted into the transmission line from an excited qubit as experimentally realized in \cite{Eichler11}. 
We here chose to operate the transistor such that the target photon is reflected in the absence but unaffected in the presence of the control photon and choose $\omega_{\text{t}}$ to be equal to the frequency of the transition $\ket{g_{1}g_{2}} \to \ket{e_{1}g_{2}}$ ($\ket{g_{j}}/\ket{e}_{j}$ denotes qubit $j$ in the ground/excited state),
i.e. $\omega_{\text{t}} = \omega_{1} + 2 J$, see figure \ref{pnse}c. The reverse mode of operation where the target photon is unaffected in the absence and reflected in the presence of the control photon can be selected by choosing $\omega_{\text{t}} = \omega_{1} - 2 J$ and works equally well.
Without control photon evidently no photon-photon interaction can take place and the output state reads,
$\ket{\Psi_{\text{out}}} = \int dp  \sum_{i=r,l,c} \beta_{i}(p) i_p^\dag \ket{0}$,
where the transmission amplitude is denoted $\beta_{r}(p)$, the reflection amplitude $\beta_{l}(p)$ and the amplitude for the target photon being lost $\beta_c(p)$. These amplitudes relate to the initial state via $\beta_{i}(p)= \int dk \, \alpha_{_{\text{t}}}(k) \, S_i(k,p)$, where the
$S_i(k,p)$ are the S-matrix elements for the different processes \cite{supplement}.
The resulting transmission probability for the target photon reads,
\begin{equation}\label{pts}
 p_T = \frac{\tau_1 + \tau_1^2 \gamma + (\tau_1 \gamma)^2 \tau_{\text{t}}}{(1+\tau_1 \gamma)(\tau_1 + \tau_{\text{t}} + \tau_{\text{t}} \tau_1 \gamma)}
\end{equation}
and the reflection probability $p_R = \tau_{\text{t}}/(1+ \tau_1 \gamma)(\tau_1 + \tau_{\text{t}}  + \tau_{\text{t}} \tau_1 \gamma)$,
where $\gamma = (\gamma_{\text{r}}/2) + \gamma_{\varphi} = 1/T_{2}^{*}$ and $T_{2}^{*}$ is the phase coherence time.
We note that $p_{T} + p_{R} < 1$ because the photon can also be lost due to qubit relaxation.
Importantly, in the regime of $T_{2}^{*} \gg \tau_{\text{t}} \gg \tau_{1}$, the reflection probability for the target photon approaches unity \cite{Chang07}. 

Next we consider the case of the same incident target photon but now in the presence of an incoming control photon. As the control photon inverts qubit 2, the scattering center is in the state $\ket{g_{1}e_{2}}$ and the target photon can only couple to the transition $\ket{g_{1}e_{2}} \to \ket{e_{1}e_{2}}$, see figure \ref{pnse}c. This transition is detuned by $4J$ from the target photon frequency, and thus the transmission probability for the target photon approaches unity as $J$ becomes larger than the linewidths of target pulse and qubit 1, $J > \tau_1^{-1} + \tau_{\text{t}}^{-1}$.
Our scheme works best if the control photon pulse is chosen such that it maximally inverts qubit 2.
A suitable pulse is thus the time reversed version of a pulse resulting from spontaneous emission of qubit 2 into the transmission line \cite{Cirac97} which is often called an inverting pulse \cite{FanInv}. The generation of inverting pulses and their release into a transmission line
was demonstrated recently \cite{Yin12}. For the cut transmission line in subsystem 2 an inverting pulse of carrier frequency $\omega_{\text{c}}$ and temporal width $\tau_{\text{c}}$ reads $\ket{\Psi_{_{\text{c}}}}= \int dk \, \alpha_{_{\text{c}}}(k) \, b_{k}^\dag \ket{0}$ with
$\alpha_{_{\text{c}}}(k) = \{\sqrt{\pi \tau_{\text{c}}} \, [- \I (\omega_{\text{c}}-k)+\tau_{\text{c}}^{-1}] \}^{-1}$.
We note that our results do not change if the target photon pulse also has the shape of an inverting pulse. Since a target pulse that is transmitted will keep its shape our scheme is thus indeed scalable.
Due to the coupling to vacuum, the qubit 2 is of course never completely inverted. 

The output state can be written as,
$\ket{\Psi_{\text{out}}} = \int dp_1 dp_2 \sum_{i=r,l,c} \sum_{j=b,d}\beta_{i,j}(p_1,p_2) i^{\dag}_{p_1} j^{\dag}_{p_2} \ket{0}$, where
the first index in the amplitudes $\beta_{i,j}(p_1,p_2)$ refers to the target photon, which can be reflected ($l$), transmitted ($r$)
or lost ($c$) and the second index refers to the control photon which can be reflected ($b$) or lost ($d$).
For the probability of the target photon being transmitted in the presence of a control photon we thus get,
\begin{equation}\label{ptsc}
p_{TC} = \int dp_1 dp_2 \left[ \left| \beta_{r,b}(p_1,p_2) \right|^2 + \left| \beta_{r,d}(p_1,p_2) \right|^2 \right] .
\end{equation}

\begin{figure}
\includegraphics[width=\columnwidth]{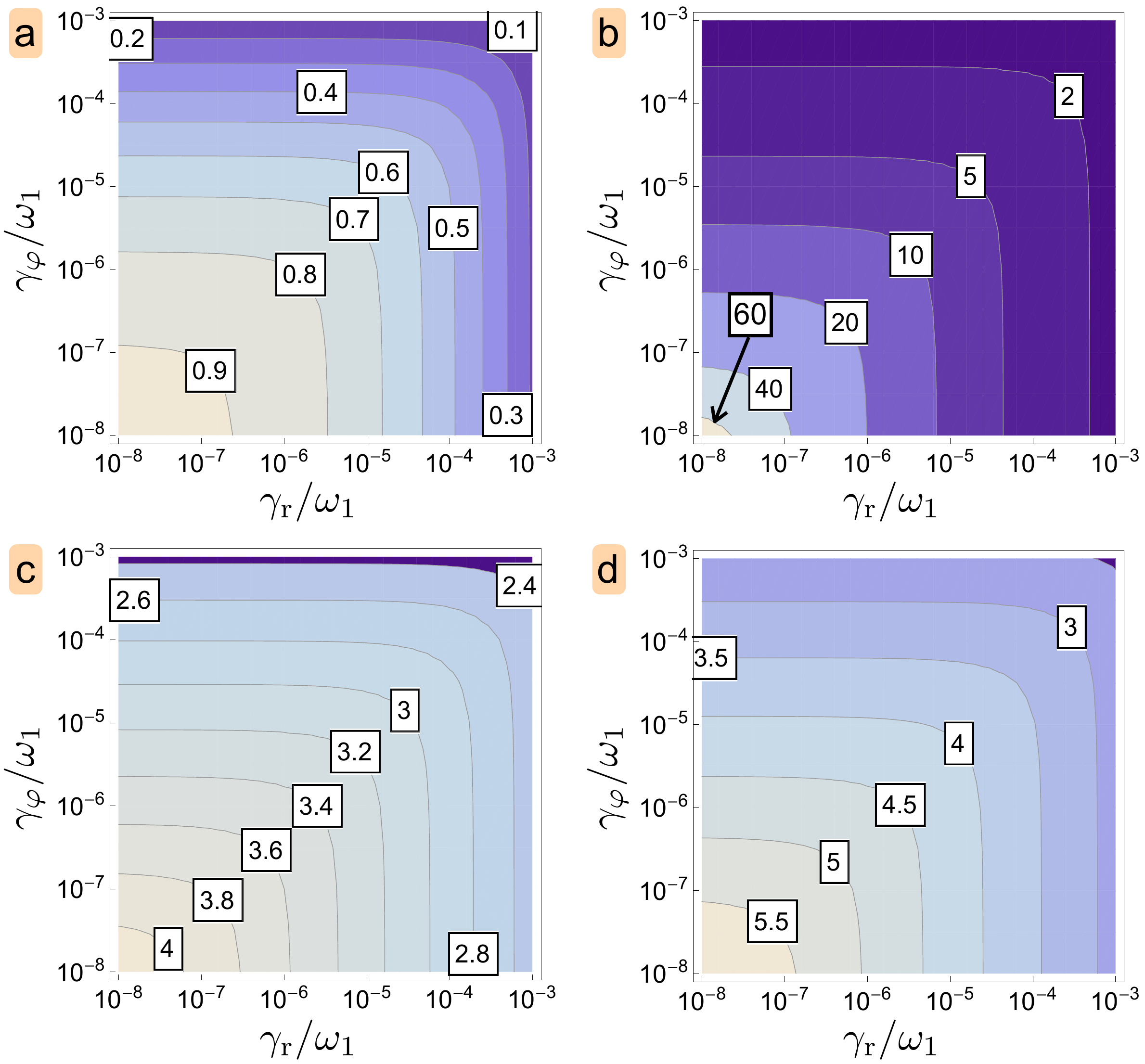}
\caption{Performance of the ``single-photon'' transistor. {\bf a:} Maximized contrast $C_s= p_{TC}-p_T$ as a function of the rates for qubit relaxation, $\gamma_{\text{r}}$, and pure dephasing, $\gamma_{\varphi}$, for $\omega_{\text{c}} = \omega_{2} + 2 J$, $\omega_{\text{t}} = \omega_{1} + 2 J$, $\omega_{2} = \omega_{1}$, $\tau_1 = 200 / \omega_1$, $J = 0.01 \, \omega_{1}$ and $\tau_{\text{c}} = \tau_{2}$. {\bf b:} On-off ratio $R_s = p_{TC}/p_T$ for the same parameters. {\bf c} and {\bf d:} $\log_{10}(\tau_{\text{t}} \omega_{1})$ and $\log_{10}(\tau_{\text{c}} \omega_{1})$ for the optimal choices of $\tau_{\text{t}}$ and $\tau_{\text{c}}$ as functions of $\gamma_{\text{r}}/\omega_1$ and $\gamma_{\varphi}/\omega_1$.}
\label{is}
\end{figure}

\begin{figure}
\centering
\includegraphics[width=0.32\textwidth]{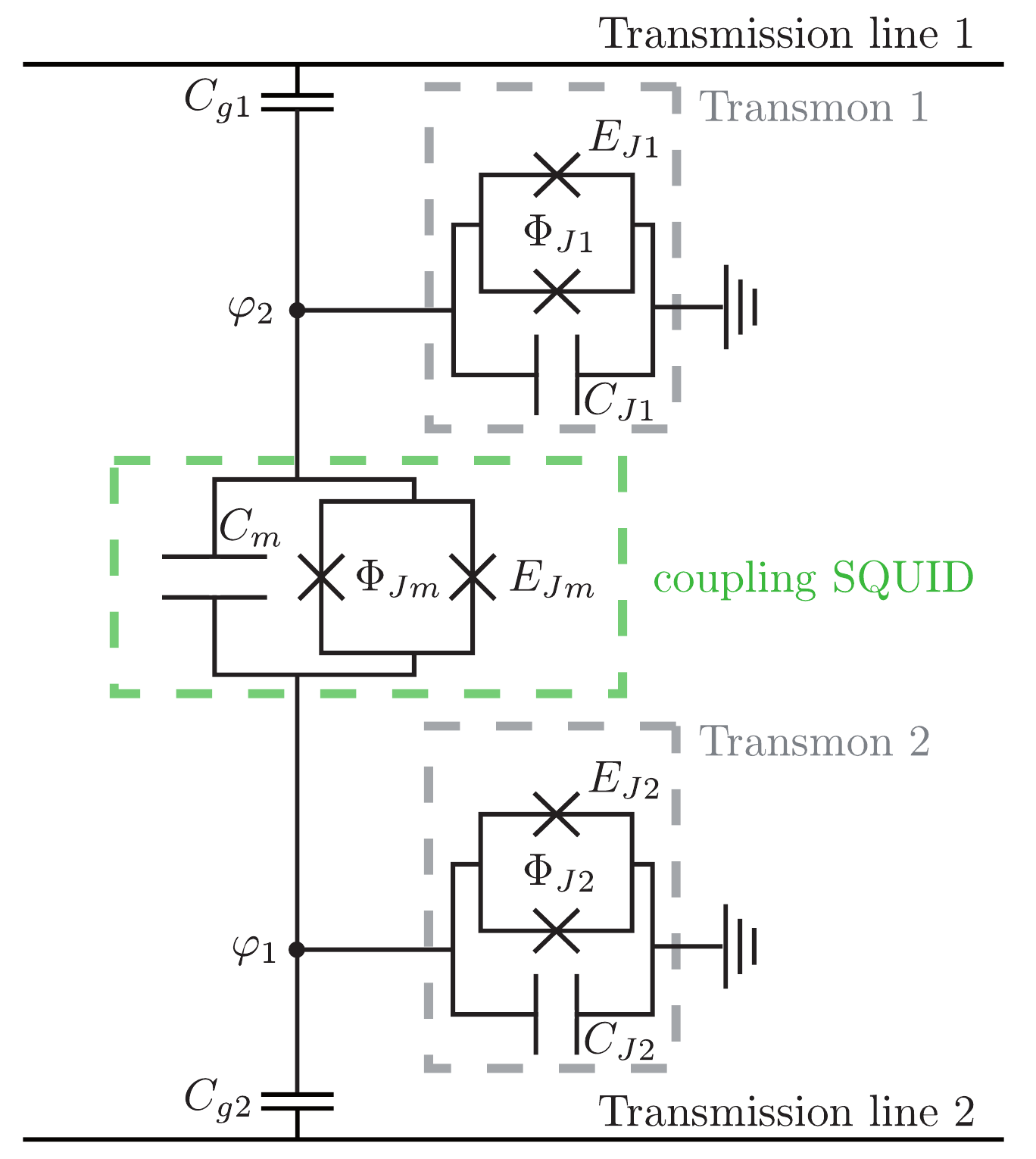}
\caption{Circuit model of the two coupled transmon qubits with Josephson energies $E_{Jj}$ and shunting capacitances $C_{j}$. Both transmons are coupled via a combination of capacitive and inductive coupling realized with a SQUID arrangement with Josephson energy $E_{Jm}$ and shunting capacitance $C_m$. The transmons are capacitively coupled to transmission lines.}
\label{transmoncircuit}
\end{figure}

We quantify the performance of the ``single-photon'' transistor we propose via the difference $C_{s}$ and ratio $R_{s}$ between the transmission probabilities for the target photon in the presence and absence of a control photon,
\begin{equation}
C_s = p_{TC}-p_T  \quad \text{and} \quad R_s = p_{TC}/p_T ,
\end{equation}
where $p_{TC}$ and $p_{T}$ are given in equations (\ref{pts}) and (\ref{ptsc}) respectively.
For $C_s = 1$ the setup would describe an ideal transistor for single photons.
Figure \ref{is} shows the achievable transmission contrast, $C_s$, and on-off ratio, $R_{s}$, for a realistic device with $\omega_{1} = \omega_{2}$ and a qubit-qubit coupling of $J = 0.01 \, \omega_{1}$ as a function of the relaxation rate $\gamma_{\text{r}}$ and pure dephasing rate $\gamma_{\varphi}$ of the qubits. 
As the plots show, an ideal ``single-photon'' transistor can be realized in the limit of vanishing $\gamma_{\text{r}}/\omega_{1}$ and $\gamma_{\varphi}/\omega_{1}$ whereas 
very good performance can already be expected for currently realized values of $J/ 2 \pi \sim  50\,$MHz, $\gamma_{\text{r}}/\omega_{1} \sim 10^{-6}$ and $\gamma_{\varphi}/\omega_{1} \sim 10^{-6}$ \cite{Fink08,Sandberg,Rigetti12}, where a single control photon changes the transmission probability for the target photon by a factor 20.
The performance of the ``single-photon'' transistor we propose depends on the shapes of the target and control photon pulses and the parameters of the Hamiltonian (\ref{Ham}).  
As expected the best choices for the carrier frequencies of the control and target pulses are equal to the transition frequencies of $\ket{g_{1}g_{2}} \to \ket{g_{1}e_{2}}$ respectively $\ket{g_{1}g_{2}} \to \ket{e_{1}g_{2}}$, i.e. $\omega_{\text{c}} = \omega_{2} + 2 J$ and $\omega_{\text{t}} = \omega_{1} + 2 J$. For a single transistor $\omega_{1}$ and $\omega_{2}$ may be chosen arbitrarily. Yet to enable concatenation of multiple transistors, we choose $\omega_{1} = \omega_{2}$. Moreover the interaction of the target photon with qubit 1 should be as high as possible. We choose $\tau_1 = 200 / \omega_1$ which is compatible with experiments. For a control photon which is an inverting pulse, the optimal choice for its temporal width is obviously $\tau_{\text{c}} = \tau_{2}$. There are thus two remaining parameters, $\tau_{\text{t}}$ and $\tau_{\text{c}}$, which we have optimized numerically. The optimal choices of $\tau_{\text{t}}$ and $\tau_{\text{c}}$ as functions of $\gamma_{\text{r}}/\omega_{1}$ and $\gamma_{\varphi}/\omega_{1}$ are shown in figures \ref{is}b and \ref{is}c respectively.

The results presented in figure \ref{is} assume that control and target pulses arrive at the same time. A possible delay between both pulses can be detrimental to the contrast $C_{s}$ and on-off ratio $R_{s}$. Yet we find that the performance of the transistor is increasingly robust against such delays with increasing phase coherence time $T_{2}^{*}$ of the qubits. For example for $T_{2}^{*} = 10^{6} \omega_{1}^{-1}$ a very good performance of the transistor is retained for delays up to $T = 10^{4} \omega_{1}^{-1}$, see supplemental material \cite{supplement} for details.

A conservative estimate for the effective 'gain' of our transistor is provided by the maximal number of target photons that can be reflected due to the presence of a single control photon. Since target photons only generate a very small excitation probability for qubit 1 and thus do not appreciably affect even ``single-photon'' control pulses, the effective 'gain' can be high. It grows with increasing phase coherence time, $T_{2}^{*}$, of the qubits and for example reaches 70 for $T_{2}^{*} = 10^{6} \omega_{1}^{-1}$, see \cite{supplement}.

Finally, for the fully scalable case where both, control and target photons are inverting pulses with the same carrier frequency, $\omega_{\text{t}} = \omega_{\text{c}}$, and pulse length, $\tau_{\text{t}} = \tau_{\text{c}}$, we find that the contrast reaches $C_{s} \approx 0.6$ for $T_{2}^{*} \ge 10^{6} \omega_{1}^{-1}$ and a control photon that arrives $\sim 4 \times 10^{3} \omega_{1}^{-1}$ ahead of the target photon, see \cite{supplement}. This strong influence of the control photon on the target photon despite their identical pulse shapes
is enabled by the asymmetry of the device with a semi-infinite (infinite) transmission line for the control (target) photon and $\tau_{1} \gg \tau_{2}$. Hence the control pulse can be matched to the control qubit with $\omega_{\text{c}} = \omega_{2}$ and $\tau_{\text{c}} = \tau_{2}$, while the target photon is not matched to its qubit.

\paragraph{Coupled transmons}

As stated above, the qubit-qubit interaction in equation (\ref{HIsing}) can be realized with two transmon qubits that are coupled via a SQUID.
The circuit we consider is sketched in figure \ref{transmoncircuit}. 
and described by the Lagrangian \cite{Yong2011}, $\mathcal{L} = \mathcal{L}_{1} + \mathcal{L}_{2} + \mathcal{L}_{12}$,
where the Lagrangians of the individual qubits read,
$\mathcal{L}_{j} = \frac{C_{j}}{2}\dot{\varphi}_j^2 + \frac{C_{gj}}{2}(\dot{\varphi}_j-V_j)^2 + E_{Jj} \cos\left(\varphi_j/\varphi_0\right)$
and
$\mathcal{L}_{12} = \frac{C_m}{2}\left(\dot{\varphi}_1-\dot{\varphi}_2\right)^2 + E_{Jm}\cos\left(\frac{\varphi_1-\varphi_2}{\varphi_0}\right)$.
Here $\varphi_0=\hbar/(2e)$ is the flux quantum divided by $2 \pi$, the $C_{j}$ and $E_{Jj}$ are the capacitances and Josephson energies of the individual transmons. $C_{m}$ and $E_{Jm}$ are the capacitance and Josephson energy of the capacitively shunted coupling SQUID. The $C_{gj}$ are the coupling capacitances between the transmission lines and the individual transmons and the $V_{j}$ are the fully quantum mechanical quadratures of the electric potential of the transmission line fields. All Josephson energies of the setup are tunable by threading external fluxes $\Phi_{Jj}$ and $\Phi_{Jm}$ through the respective SQUID loops, c.f. figure \ref{transmoncircuit}.
We write the corresponding Hamiltonian of the transmons in terms of creation and annihilation operators $a_{j}^{\dagger}$ and $a_{j}$ \cite{supplement}.
By tuning the $E_{Jj}$ and $E_{Jm}$ such that $\frac{E_{Jm}}{\sqrt{E_{J1} + E_{Jm}}\sqrt{E_{J2}+E_{Jm}}}=\frac{C_m}{\sqrt{C_1+C_m}\sqrt{C_2+C_m}}
$, all interactions of the form $a_1a_2^{\dag}+a_1^{\dag}a_2$
cancel and the leading term of the remaining interactions reads $-2\frac{E_{Jm} \sqrt{E_{C1}E_{C2}}}{\sqrt{E_{J1} + E_{Jm}}\sqrt{E_{J2}+E_{Jm}}} a_1^{\dag}a_1a_2^{\dag}a_2$ which is equivalent to the interaction in equation (\ref{HIsing}) with $J= 2\frac{E_{Jm} \sqrt{E_{C1}E_{C2}}}{\hbar \sqrt{E_{J1} + E_{Jm}}\sqrt{E_{J2}+E_{Jm}}}$. Within the approximations we use \cite{supplement} the achievable qubit-qubit coupling is $J < \sqrt{E_{C1}E_{C2}} / (10 \hbar )$.

In conclusion, we have introduced a scheme for a ``single-photon'' transistor in circuit quantum electrodynamics
that is inherently scalable as both photons can have the same carrier frequency and pulse shape. Due to its 'gain' the device could also detect single photons that propagate in the control line. Moreover it works quantum coherently such that e.g. a control pulse consisting of a superposition of a single photon and the vacuum generates a quantum superposition of the target photon being blocked and transmitted. Its performance might be further improved by suppressing losses with multiple, regularly spaced qubit pairs \cite{Zoubi10}.
Moreover, the complexity of a network built with such transistors, c.f. figure \ref{pnse}b, could be increased further by integrating directional couplers between
them \cite{Ku11}.

\acknowledgements
\paragraph{Acknowledgements}
This work is part of the Emmy Noether project HA 5593/1-1 and the CRC 631, both funded by the German Research Foundation, DFG.

\pagebreak
\widetext

\begin{center}
\large
\bf
Supplemental Material
\end{center}

\appendix
\section{Langevin equations} \label{sec:langevin}

Here we present the explicit forms of the Langevin equations \cite{GardinerZoller} for photon and qubit operators that describe the unitary dynamics generated by the Hamiltonian $H$ in equation (3) of the main text and the dissipative processes associated to qubit relaxtion and pure dephasing at rates $\gamma_{\text{r}}$ respectively $\gamma_{\varphi}$. The equations read,
\begin{eqnarray}
\dot{\sigma}_{1}^{-} & = & - \left(\I \omega_{1} + \frac{1}{\tau_{1}} + \frac{\gamma_{\text{r}}}{2} + \gamma_{\varphi}\right) \sigma_{1}^{-} - 2 \I J \sigma_{2}^{z} \sigma_{1}^{-} + \I \sqrt{\frac{2}{\tau_{1}}} \sigma_{1}^{z} a_{\text{in}} + \I \sqrt{\gamma_{\text{r}}} \sigma_{1}^{z} c_{\text{in}} -\I \sqrt{2 \gamma_{\varphi}} (\sigma_{1}^{-} \tilde{c}_{\text{in}} + \tilde{c}_{\text{in}}^{\dag} \sigma_{1}^{-})\\
\dot{\sigma}_{1}^{z} & = & - \left(\frac{2}{\tau_{1}} + \gamma_{\text{r}} \right) (\sigma_{1}^{z} + \id) + 2 \I \sqrt{\frac{2}{\tau_{1}}} (a_{\text{in}}^{\dag} \sigma_{1}^{-} - \sigma_{1}^{+} a_{\text{in}}) + 2 \I \sqrt{\gamma_{\text{r}}} (c_{\text{in}}^{\dag} \sigma_{1}^{-} - \sigma_{1}^{+} c_{\text{in}})\\
\dot{\sigma}_{2}^{-} & = & - \left(\I \omega_{2} + \frac{1}{\tau_{2}} + \frac{\gamma_{\text{r}}}{2} + \gamma_{\varphi}\right) \sigma_{2}^{-} - 2 \I J \sigma_{1}^{z} \sigma_{2}^{-} + \I \sqrt{\frac{2}{\tau_{2}}} \sigma_{2}^{z} b_{\text{in}} + \I \sqrt{\gamma_{\text{r}}} \sigma_{2}^{z} d_{\text{in}} -\I \sqrt{2 \gamma_{\varphi}} (\sigma_{2}^{-} \tilde{d}_{\text{in}} + \tilde{d}_{\text{in}}^{\dag} \sigma_{2}^{-})\\
\dot{\sigma}_{2}^{z} & = & - \left(\frac{2}{\tau_{2}} + \gamma_{\text{r}} \right) (\sigma_{2}^{z} + \id) + 2 \I \sqrt{\frac{2}{\tau_{2}}} (b_{\text{in}}^{\dag} \sigma_{2}^{-} - \sigma_{2}^{+} b_{\text{in}}) + 2 \I \sqrt{\gamma_{\text{r}}} (d_{\text{in}}^{\dag} \sigma_{2}^{-} - \sigma_{2}^{+} d_{\text{in}}),
\end{eqnarray}
where $a_{\text{in}} = (r_{\text{in}} + l_{\text{in}}) / \sqrt{2}$. Moreover $c_{\text{in}}$ ($d_{\text{in}}$) are the noise operators associated to the relaxation of qubit 1 (qubit 2) and $\tilde{c}_{\text{in}}$ ($\tilde{d}_{\text{in}}$) are the noise operators associated to the pure dephasing of qubit 1 (qubit 2).
For a derivation see e.g. \cite{Blencowe2008}. From these equations, the desired output fields $a_{\text{out}}$ ($b_{\text{out}}$) for transmission line 1 (2) can be computed via the input-output relations \cite{Gardiner},
\begin{equation}\label{inout4}
 a_{\text{out}}(t) =  a_{\text{in}}(t) - \I \sqrt{\frac{2}{\tau_{1}}} \sigma_{1}^{-}(t) \quad \text{and} \quad
 b_{\text{out}}(t) =  b_{\text{in}}(t) - \I \sqrt{\frac{2}{\tau_{2}}} \sigma_{2}^{-}(t)
\end{equation}

\section{Scattering theory and input-output relations}

We calculate the output states of our device for given input states by making use of a combination of scattering theory and the input-output formalism. In this approach, output and input states are connected via the scattering matrix (S-matrix),
\begin{equation}\label{defS}
\ket{\Psi_{\text{out}}} = S \ket{\Psi_{\text{in}}}
\end{equation}
The S-matrix elements in turn can be written in terms of scattering operators \cite{taylor}.
For example for the process of transmission of a single photon, the S-matrix element reads,
\begin{equation}\label{onep}
 S_r(k,p)= \bra{0} r_{\text{out}}(p) r^\dag_{\text{in}}(k) \ket{0},
\end{equation}
where
\begin{equation}\label{bak1}
r_{\text{in}/\text{out}}^{\dag}(\omega) =  \lim_{t_0\to \mp\infty} e^{\I H t_0} e^{-\I H_T t_0} r^{\dag}(\omega) e^{\I H_T t_0} e^{-\I H t_0},
\end{equation}
and $H_T = \int^{\infty}_{-\infty} dp \, p \, (r_p^{\dagger} r_p - l_p^{\dagger} l_p) + \int^{\infty}_{-\infty} dp \, p \, b_p^{\dagger} b_p$ is as in the main text.
In turn, for a two photon process, where the target photon is transmitted and the control photon is lost into the vacuum, the S-matrix element reads,
\begin{equation}\label{twop}
S_{r,d}(k,k',p,p') = \bra{0} r_{\text{out},p} d_{\text{out},p'} r^\dag_{\text{in},k} b^\dag_{\text{in},k'} \ket{0}
\end{equation}
For our system it is convenient to introduce even and odd modes for transmission line 1,
\begin{equation}
a(p) = \frac{1}{\sqrt{2}} \left( r_{p} + l_{-p} \right) \quad \text{and} \quad
\mathring{a}(p) = \frac{1}{\sqrt{2}} \left( r_{p} - l_{-p} \right) \label{evenodd}
\end{equation}
Due to the form of the photon-qubit coupling in equation (2) of the main text, only even modes couple to the qubit whereas odd modes completely decouple from the rest of the system and describe freely propagating photons.
The coupling strength to the even modes is enhanced by a factor of $\sqrt{2}$.

To calculate the desired scattering matrix elements we exploit a link to the equations of motion, i.e. the Langevin equations in section \ref{sec:langevin}, that describe the dynamics of the scattering center.
To make use of the Langevin equations, it is necessary to connect the scattering operators to standard input-output operators that fulfill input-output relations, which e.g. for transmission line 1 read,
\begin{equation}\label{inout3}
 a_{\text{out}}(t) =  a_{\text{in}}(t) - \I \sqrt{\frac{2}{\tau_{1}}} \sigma_{1}^{-}(t)
\end{equation}
where $a_{\text{in}}(t)$ and $a_{\text{out}}(t)$ are any even mode input/output operators, defined by
\begin{equation}\label{g10}
 a_{\text{in}/\text{out}}(t) \equiv \frac{1}{\sqrt{2 \pi}} \int d\omega e^{- \I \omega (t-t_{0/1})}a_{0/1}(\omega).
\end{equation}
Here, the operator $a_{0/1}(\omega)$ plays the role of an initial value in the Heisenberg picture,
\begin{equation}\label{hrhr}
 a_{0/1}(\omega) = e^{\I H t_{0/1}} a(\omega) e^{-\I H t_{0/1}}
\end{equation}
The corresponding input and output scattering operators for even modes are defined as in equation (\ref{bak1}),
\begin{equation}\label{bak}
a_{\text{in}/\text{out}}(\omega) =  \lim_{t_0\to \mp\infty} e^{\I H t_0} e^{-\I H_T t_0} a(\omega) e^{\I H_T t_0} e^{-\I H t_0} =\lim_{t_0\to \mp\infty} e^{\I \omega t_0} e^{\I H t_0} a(\omega) e^{-\I H t_0},
\end{equation}
where we have made use of the Baker-Campbell-Hausdorff formula.
We now solve equation (\ref{bak}) for $a(\omega)$ and plug the result into equation (\ref{g10}) by making use of equation (\ref{hrhr}).
Taking the limits $t_{0} \to -\infty$ respectively $t_{1} \to +\infty$ in equation (\ref{g10}) 
we find that the scattering operator $a_{\text{in}}(\omega)$ is the Fourier transform of the Heisenberg operator $a_{\text{in}}(t)$ in the limit $t_0 \to - \infty$ and that the the scattering operator $a_{\text{out}}(\omega)$ is the Fourier transform of the Heisenberg operator $a_{\text{out}}(t)$ in the limit $t_1 \to + \infty$ \cite{Fan},
\begin{equation}\label{FT}
a_{\text{in}/\text{out}}(t) = \frac{1}{\sqrt{2 \pi}} \int d\omega a_{\text{in}/\text{out}}(\omega) e^{-\I \omega t}
\end{equation}
Using equation (\ref{FT}) we express the scattering matrix $S$ in terms of input and output operators which are found from solutions of the Langevin equations in section \ref{sec:langevin}, that describe the dynamics of the scattering center.

\section{Time delays between control and target pulses}
\label{sec:delays}

In this section we present explicit estimates for the performance of our device in cases where the pulses of the control and target photon reach the transistor at different times. As the calculations for a finite time delay between the control and target pulses are numerically rather cumbersome, we obtained these estimates with a phenomenological model that shows excellent agreement with the exact calculations in the main text for zero time delay between both pulses.

The working principle of our device is that the control pulse excites qubit 2 which in turn shifts the transition frequency of qubit 1 by $4 J$ such that the target photon is either reflected or transmitted.
In our phenomenological description we thus consider a reduced model where the target photon propagates in transmission line 1 which couples to an individual effective qubit, the transition frequency of which is $\tilde{\omega}$ for cases where qubit 2 is in its ground  state and $\tilde{\omega} + 4 J$ for case where qubit 2 is excited. We compute the transmission probability for the target photon as a function of the transition frequency of this effective qubit, $\tilde{p}_{T}(\omega)$. The total transmission probability for the target photon is then
\begin{equation}
p_{T} \approx (1 - p_{e}) \, \tilde{p}_{T}(\tilde{\omega}) + p_{e} \, \tilde{p}_{T}(\tilde{\omega}+4J).
\end{equation}
Here $p_{e}$ is the probability for the effective qubit to have a transition frequency $\tilde{\omega}+4J$ which is equal to the probability for qubit 2 to be excited and the target photon to be present at qubit 1  at the same time, i.e. 
\begin{equation}
p_{e} = \int_{-\infty}^{\infty} dt \, |\alpha_{\text{t}}(t)|^{2} P_{ex}(t+T),
\end{equation}
where $P_{ex}(t)$ is the probability that a control pulse arriving at time $t$ excites qubit 2 and $T$ denotes a possible delay between the arrival times of the control and target pulses. The time integral here represents the fact that the absolute arrival time of the photons is irrelevant and only a possible time delay between control and target photon or vice versa matters. For the control pulse considered in the main text, $P_{ex}(t)$ reads,
\begin{equation}
P_{ex}(t) = \frac{4 \tau_{\text{c}} \tau_{2}}{(\tau_{\text{c}} + \tau_{2} + \gamma \tau_{\text{c}} \tau_{2})^2}
\left\{
\begin{array}{lcr}
e^{t \left(2\gamma+\frac{2}{\tau_{2}}\right)} & \text{for} & t < 0 \\
e^{-t \left(2 \gamma +\frac{2}{\tau_{2}}\right)} & \text{for} & t \ge 0 ,
\end{array}
\right.
\end{equation}
where $\gamma = (\gamma_{\text{r}}/2) + \gamma_{\varphi} = 1/T_{2}^{*}$.
We have assumed that $p_{e}$ is independent of the presence of the target photon which is a good assumption as the target pulse only causes negligible excitation of qubit 1. 

\begin{figure}[h]
\centering
\includegraphics[width=14cm]{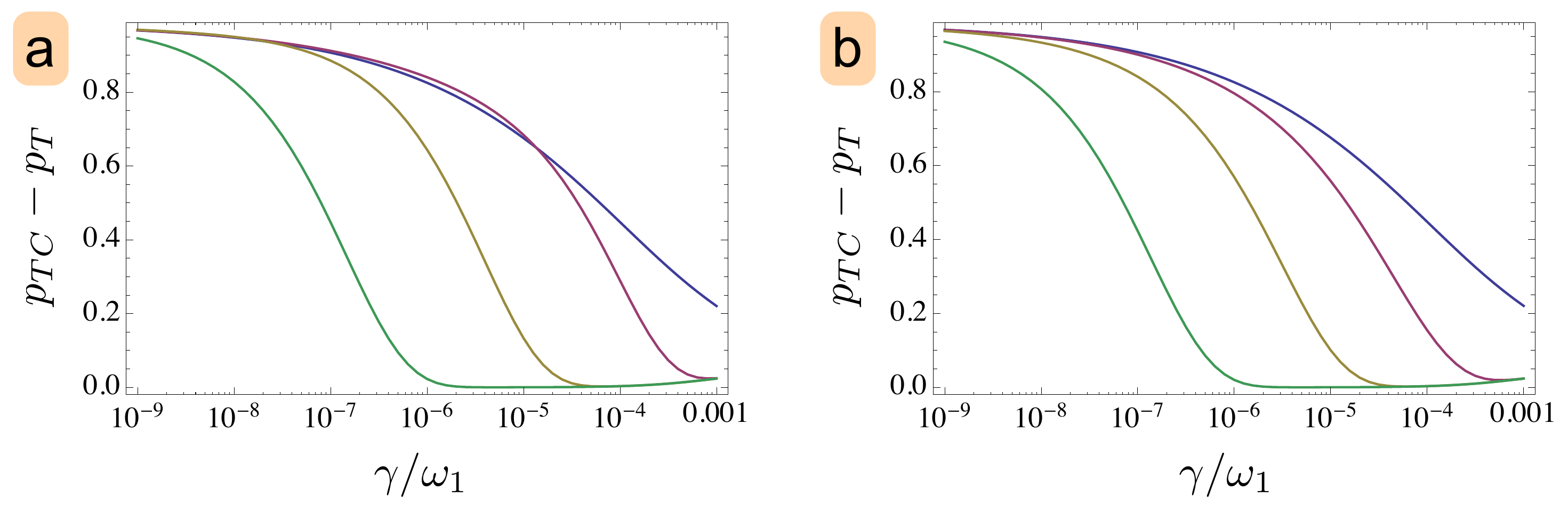}
\caption{Maximized contrast $C_s= p_{TC}-p_T$ as a function of $\gamma/\omega_1$, where $\gamma = (\gamma_{\text{r}}/2) + \gamma_{\varphi} = 1/T_{2}^{*}$, for various time delays between the arrival of control and target pulses.  {\bf a:} Scenario where the control pulse arrives before the target pulse for delays $T = 0 \times \omega_{1}^{-1}$ (blue), $T = 10^{3} \times \omega_{1}^{-1}$ (magenta), $T = 10^{4} \times \omega_{1}^{-1}$ (yellow) and $T = 10^{5} \times \omega_{1}^{-1}$ (green). {\bf b:} Scenario where the control pulse arrives after the target pulse for delays $T = 0 \times \omega_{1}^{-1}$ (blue), $T = - 10^{3} \times \omega_{1}^{-1}$ (magenta), $T = - 10^{4} \times \omega_{1}^{-1}$ (yellow) and $T = - 10^{5} \times \omega_{1}^{-1}$ (green).  The remaining parameters are as in figure 2 of the main text, $\omega_{\text{c}} = \omega_{2} + 2 J$, $\omega_{\text{t}} = \omega_{1} + 2 J$, $\omega_{2} = \omega_{1}$, $\tau_1 = 200 / \omega_1$, $J = 0.01 \, \omega_{1}$ and $\tau_{\text{c}} = \tau_{2}$.}
\label{delays1}
\end{figure}

Figure \ref{delays1} shows the maximized contrast $C_s= p_{TC}-p_T$ as a function of $\gamma/\omega_1$ for various time delays between the arrival of control and target pulses.
Interestingly, it can even be beneficial that the control photon arrives before the target photon, see also section \ref{sec:scalable}.

\section{Effective gain of the transistor}

\begin{figure}[h]
\centering
\includegraphics[width=7cm]{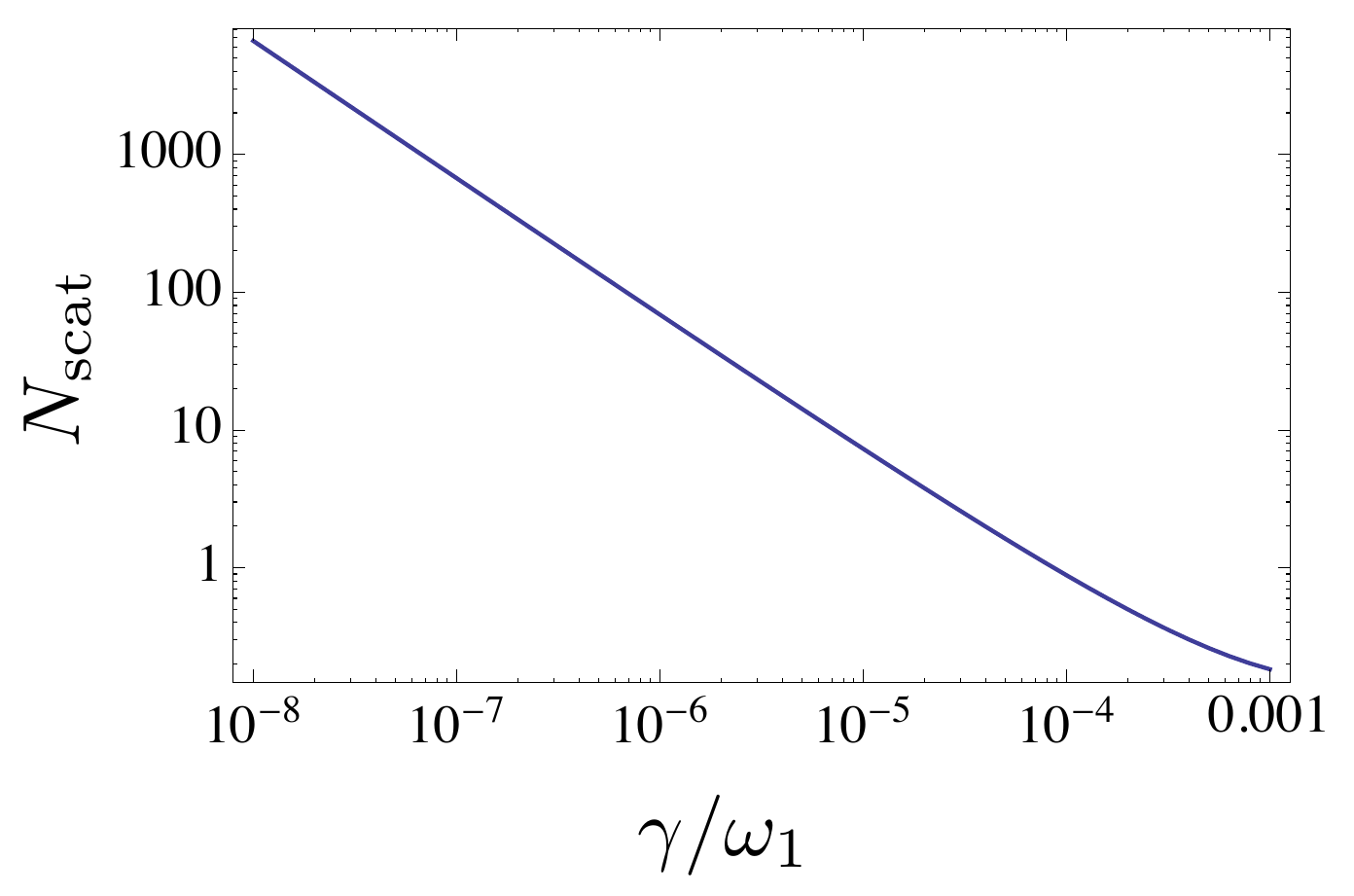}
\caption{Maximal number of target photons $N_{\text{scat}}$ that can be reflected at the qubits due to the presence of one control photon, where the probability for perturbing the control photon that is at most 5\%, as a function of $\gamma/\omega_1$, where $\gamma = (\gamma_{\text{r}}/2) + \gamma_{\varphi} = 1/T_{2}^{*}$. The remaining parameters are as in figure 2 of the main text, $\omega_{\text{c}} = \omega_{2} + 2 J$, $\omega_{\text{t}} = \omega_{1} + 2 J$, $\omega_{2} = \omega_{1}$, $\tau_1 = 200 / \omega_1$, $J = 0.01 \, \omega_{1}$ and $\tau_{\text{c}} = \tau_{2}$.}
\label{gain}
\end{figure}

To provide a conservative estimate for the effective gain of our transistor, we calculate the number of photons in the target pulse that can be reflected at the qubits due to the presence of a single control photon without having an appreciable back-action onto the control photon \cite{Chang07}. The transistor can have a large gain since the target photons only generate a very small excitation probability for qubit 1 and a large number of photons can therefore be reflected without appreciably affecting a single photon control pulse at qubit 2. The achievable gain grows with increasing phase coherence time $T_{2}^{*}$ of the qubits since a longer phase coherence time allows to use target pulses of larger temporal and thus smaller spectral width which result in a smaller probability for qubit 1 to be excited and hence a smaller back-action onto the control pulse. Figure \ref{gain} shows the maximal number of target photons that can be reflected while keeping the probability for a perturbation of the control photon below 5\%. For example for $T_{2}^{*} = 10^{6} \omega_{1}^{-1}$ one can scatter 70 target photons with a single control photon.
The estimate for $N_{\text{scat}}$ is obtained with the same approximation as described in section \ref{sec:delays}, but where the roles of target and control photons are interchanged.
We furthermore assume that the target photons scatter independently of each other at qubit 1. This assumption is very accurate for our setting since qubit 1 is mostly in its ground state and has a vanishingly small excitation probability. 

We note that the above estimate provides a lower bound for the gain of our transistor. Indeed, the gain may be considerably larger in the complementary mode of operation, where the target photons are transmitted in the presence of a control photon but blocked in its absence, because the requirement that the control photon remains unperturbed is no longer necessary for that case.  Moreover, in the absence of the control photon, an arbitrarily large number of target photons can be transmitted without interacting with the qubits as they are sufficiently detuned from the transition frequency of qubit 1.

\section{Identical Control and target pulses: the fully scalable case}

\label{sec:scalable}

\begin{figure}[h]
\centering
\includegraphics[width=7cm]{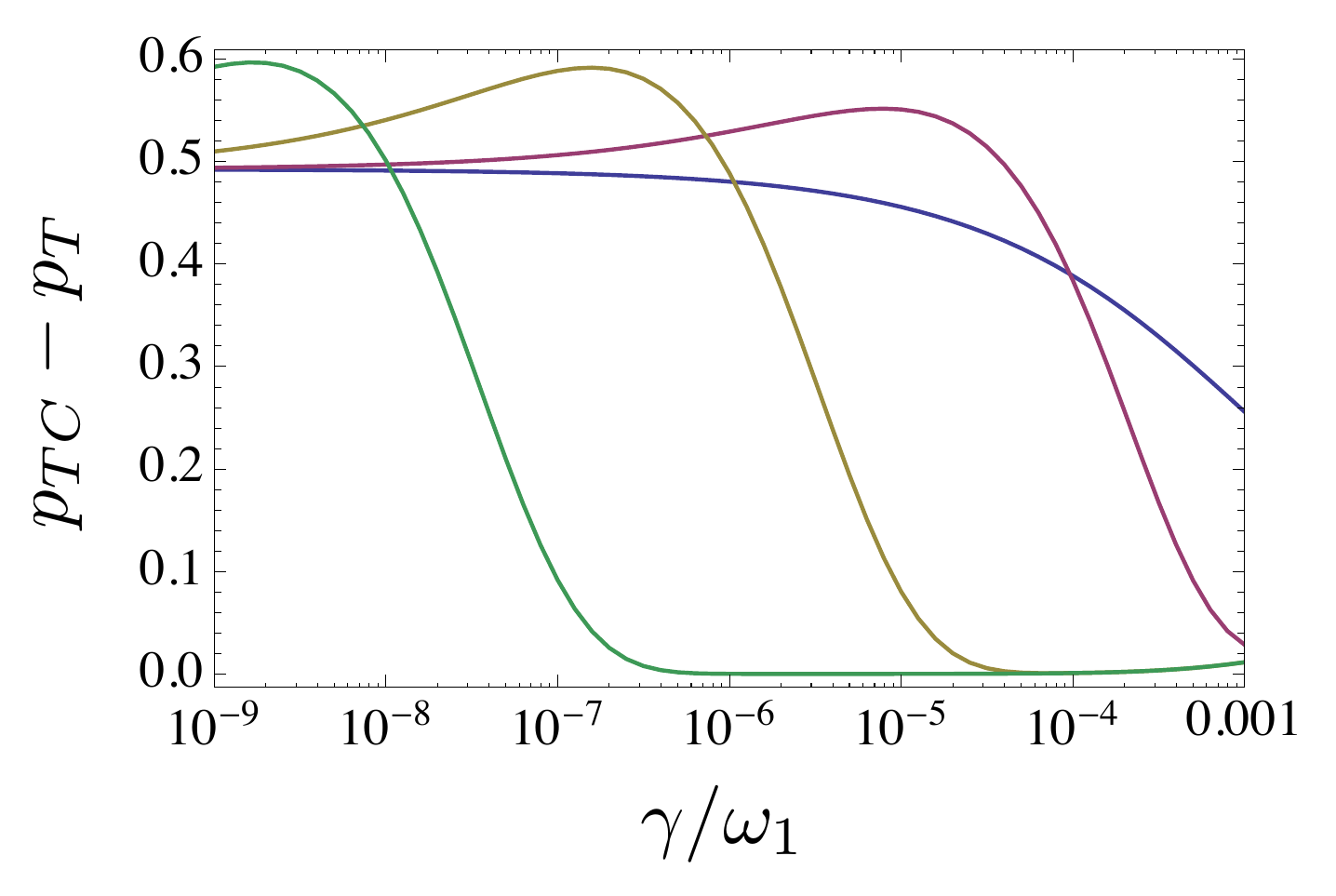}
\caption{Maximized contrast $C_s= p_{TC}-p_T$ as a function of $\gamma/\omega_1$, where $\gamma = (\gamma_{\text{r}}/2) + \gamma_{\varphi} = 1/T_{2}^{*}$, for optimized pulse lengths $\tau_{\text{c}} = \tau_{\text{t}}$ and various time delays between the arrival of control and target pulses in cases where the control and target pulses have identical shapes. Scenario where the control pulse arrives before the target pulse for delays $T = 0 \times \omega_{1}^{-1}$ (blue), $T = 10^{3} \times \omega_{1}^{-1}$ (magenta), $T = 10^{4} \times \omega_{1}^{-1}$ (yellow) and $T = 10^{5} \times \omega_{1}^{-1}$ (green). The remaining parameters are as in figure 2 of the main text, $\omega_{\text{c}} = \omega_{2} + 2 J$, $\omega_{\text{t}} = \omega_{1} + 2 J$, $\omega_{2} = \omega_{1}$, $\tau_1 = 200 / \omega_1$, $J = 0.01 \, \omega_{1}$ and $\tau_{\text{c}} = \tau_{2}$.}
\label{scalable1}
\end{figure}

\begin{figure}[h]
\centering
\includegraphics[width=\textwidth]{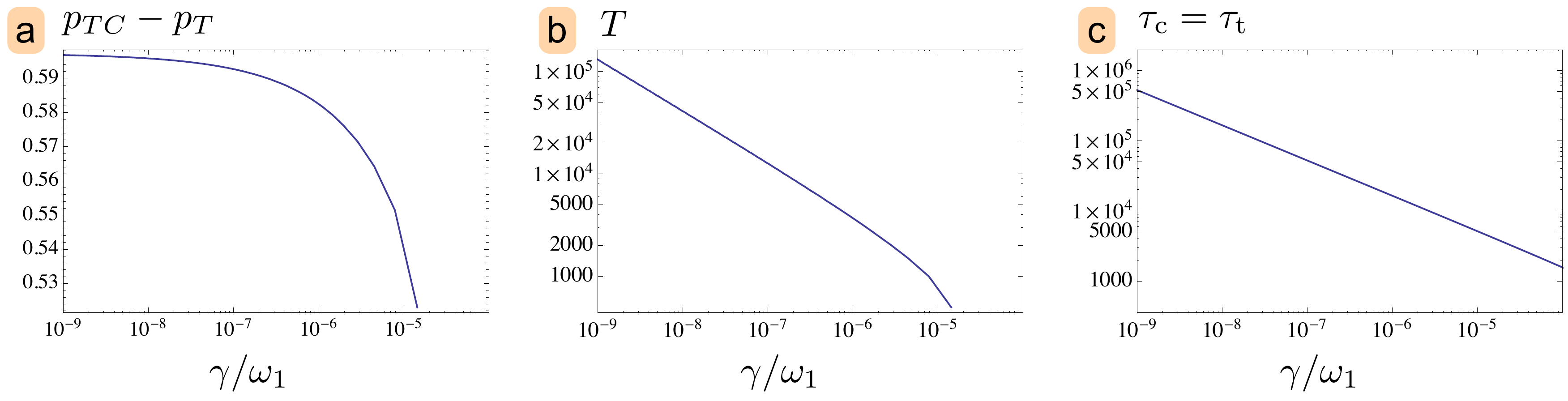}
\caption{{\bf a:} Maximized contrast $C_s= p_{TC}-p_T$ as a function of $\gamma/\omega_1$ for optimized pulse length $\tau_{\text{c}} = \tau_{\text{t}}$ and delay $T$. {\bf b:} Optimal choice for the delay $T$ between control and target pulse. {\bf c:} Optimal choice for the lengths $\tau_{\text{c}} = \tau_{\text{t}}$ of both, control and target pulses. The remaining parameters are as in figure 2 of the main text, $\omega_{\text{c}} = \omega_{2} + 2 J$, $\omega_{\text{t}} = \omega_{1} + 2 J$, $\omega_{2} = \omega_{1}$, $\tau_1 = 200 / \omega_1$, $J = 0.01 \, \omega_{1}$ and $\tau_{\text{c}} = \tau_{2}$.}
\label{scalable2}
\end{figure}

In this section we present explicit estimates for the performance of our device for the case where the pulses of the control and target photon have identical shapes. This case is of particular interest for concatenating multiple transistors in a network as an outgoing target photon of one transistor can act as a control photon for a consecutive transistor. 
The estimates in this section are again obtained with the phenomenological model introduced in section \ref{sec:delays} of this supplementary material.
Figure \ref{scalable1} shows the maximized contrast $C_s= p_{TC}-p_T$ as a function of $\gamma/\omega_1$ for control and target pulses with identical shapes, where the pulse lengths $\tau_{\text{c}} = \tau_{\text{t}}$ have been optimized.
For a given coherence time $T_{2}^{*} = 1/\gamma$ there is thus an optimal choice for the delay $T$ between control and target photon and for the lengths $\tau_{\text{c}} = \tau_{\text{t}}$ of both pulses. In figure \ref{scalable2} we plot the achievable contrast $C_s= p_{TC}-p_T$ together with the optimal values for the delay $T$ between both pulses and their lengths $\tau_{\text{c}} = \tau_{\text{t}}$ as functions of the inverse coherence time $\gamma$.
We find that the maximal contrast is rather insensitive to variations in the pulse length. It furthermore becomes less sensitive to imprecisions in the timing of the pulses with increasing phase coherence time $T_{2}^{*} = 1/\gamma$ of the qubits. As the optimal delay increases with $T_{2}^{*}$ we find that a relative precision of 50\% for the timing of the pulses would be sufficient to reach the optimal contrast with good accuracy.

\section{Coupled Transmons}

\begin{figure}[h]
\includegraphics[width=0.4\textwidth]{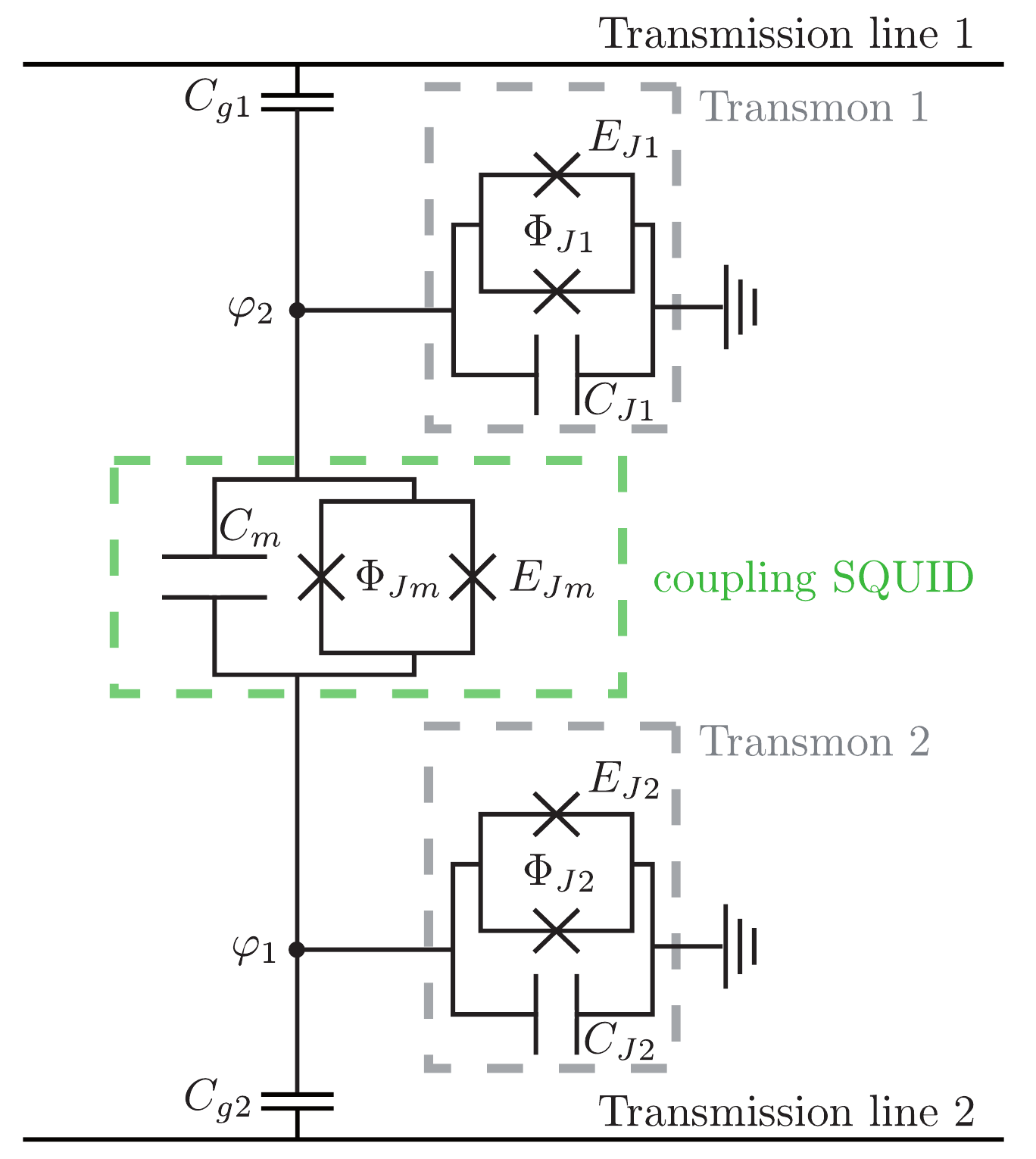}
\caption{Circuit model of the pair of transmons with Josephson energy $E_{Ji}$ and shunting capacitance $C_{Ji}$. Both transmons are coupled via a combination of capacitive and inductive coupling realized with a SQUID arrangement with Josephson energy $E_{Jm}$ and shunting capacitance $C_m$. The transmons are capacitively coupled to transmission lines and all Josephson energies of the setup are tunable by threading external fluxes $\Phi_{Ji}$ and $\Phi_{Jm}$ through respective SQUID loops.}
\label{couplingModel}
\end{figure}

The Lagrangian describing the two transmons that are mutually coupled via a SQUID and each couple to a transmission line reads,
\begin{eqnarray}
\mathcal{L}&=&\frac{C_{J1}}{2}\dot{\varphi}_1^2+\frac{C_{J2}}{2}\dot{\varphi}_2^2+\frac{C_m}{2}\left(\dot{\varphi}_1-\dot{\varphi}_2\right)^2 +\frac{C_{g1}}{2}(\dot{\varphi}_1-V_1)^2+\frac{C_{g2}}{2}(\dot{\varphi}_2-V_2)^2 \\
&+&E_{J1}\cos\left(\frac{\varphi_1}{\varphi_0}\right)+E_{J2}\cos\left(\frac{\varphi_2}{\varphi_0}\right) +E_{Jm}\cos\left(\frac{\varphi_1-\varphi_2}{\varphi_0}\right) \nn
\end{eqnarray}
where $\varphi_0=\hbar/(2e)$ is the flux quantum divided by $2\pi$, $C_{Ji}$ and $E_{Ji}$ are the capacitance and Josephson energy of the individual transmons. $C_{m}$ and $E_{Jm}$ are the capacitance and Josephson energy of the capacitively shunted coupling SQUID. The $C_{gi}$ are the coupling capacitances between the transmission lines and the individual transmons and the $V_{i}$ are the fully quantum mechanical electric potential quadratures of the transmission line fields, c.f. figure \ref{couplingModel}. 
Hence, the corresponding Hamiltonian reads,
\begin{eqnarray}
H&=&4 E_{C1}n_1^2+4 E_{C2}n_2^2+8E_{C_m} n_1n_2\\
&+&8E_{C1} n_{g,1}n_1+ 8 E_{C2} n_{g,2} n_2+ 8E_{C_m}\left(n_{g,1}n_2+n_{g,2}n_1\right) \nn \\
&-&E_{J1} \cos\phi_1 - E_{J2} \cos\phi_2-E_{Jm}\cos(\phi_1-\phi_2), \nn
\end{eqnarray}
where $\phi_{i} = \varphi_{i}/\varphi_{0}$, $n_{g,i}=\varphi_0C_{gi}V_i$,
$E_{Ci}= \frac{\hbar^{2}}{8 \varphi_{0}^{2}} \frac{C_i+C_m}{C_2C_m+C_1(C_2+C_m)}$ and $E_{C_m}=\frac{\hbar^{2}}{8 \varphi_{0}^{2}} \frac{C_m}{C_2C_m+C_1(C_2+C_m)}$ with $C_i=C_{Ji}+C_{gi}$.
The secondary capacitive coupling between transmission line 1 and transmon 2 and vice versa is small compared to the other couplings provided that $C_{m} \ll (C_{g2}/C_{g1}) C_{1}$ and $C_{m} \ll (C_{g1}/C_{g2}) C_{2}$. We thus neglect such couplings and obtain for the coupling between the transmission lines and the transmons,
\begin{equation}
H_g=8E_{C1}n_{g,1}n_1+8E_{C2}n_{g,2}n_2
\end{equation}
We furthermore separate the Josephson terms into local and nonlocal terms,
\begin{equation}
E_{Jm}\cos(\phi_1-\phi_2)=E_{Jm}\underbrace{\left(\cos\phi_1+\cos\phi_2\right)}_{\text{local}}+E_{Jm}\underbrace{\left[(\cos\phi_1-1)(\cos\phi_2 -1)+\sin\phi_1 \sin\phi_2\right]}_{\text{nonlocal}},
\end{equation}
where we have added an irrelevant constant. This leaves us with the local Hamiltonians for each transmon ($i = 1,2$),
\begin{equation}
H_i=4E_{Ci} n_i^2-\overline{E}_{Ji} \cos\phi_i,
\end{equation}
where $\overline{E}_{Ji} = E_{Ji}+E_{Jm}$ and the coupling Hamiltonian,
\begin{equation}
H_{12}=8E_{C_m}n_1n_2 - E_{Jm}\left[(\cos\phi_1-1)(\cos\phi_2 -1)+\sin\phi_1 \sin\phi_2\right]
\end{equation}
We describe the transmons in the approved approximation with anharmonic oscillators \cite{Koch07}
and introduce raising and lowering operators $a_{i}^{\dag}$ and $a_{i}$ via
\begin{eqnarray}
n_i&=&\frac{\I}{2}\left(\frac{\overline{E}_{Ji}}{2E_{Ci}}\right)^{\frac{1}{4}}(a_i-a_i^{\dag})\\
\phi_i&=&\left(\frac{2E_{Ci}}{\overline{E}_{Ji}}\right)^{\frac{1}{4}}(a_i+a_i^{\dag})
\end{eqnarray}
Keeping only the leading nonlinear terms we thus find,
\begin{equation}
H_i\approx \sqrt{8E_{Ci}\overline{E}_{Ji}} \, a_i^{\dag}a_i-\frac{E_{Ci}}{2}\, a_i^{\dag}a_i^{\dag}a_ia_i
\end{equation}
We are interested in a scenario where tunneling of excitations from one transmon to the other is strongly suppressed.
The leading tunneling terms are found by expanding the coupling Hamiltonian $H_{12}$ to linear order in $a_{1}$ and $a_{2}$. 
With a rotating wave approximation we find,
\begin{equation}\label{rwatunnel}
8E_{C_m}n_1n_2-E_{Jm}\phi_1\phi_2 \approx  \sqrt{2}(\overline{E}_{J1}\overline{E}_{J2})^{\frac{1}{4}}(E_{C1}E_{C2})^{\frac{1}{4}}\left(\frac{E_{C_m}}{\sqrt{E_{C1} E_{C2}}}-\frac{E_{Jm}}{\sqrt{\overline{E}_{J1} \overline{E}_{J2}}}\right) (a_1a_2^{\dag}+a_1^{\dag}a_2) 
\end{equation}
These terms vanish if one chooses the external fluxes that control the values of $E_{Ji}$ and $E_{Jm}$ such that,
\begin{equation} \label{eq:notunnelcond}
\frac{E_{Jm}}{\sqrt{E_{J1} + E_{Jm}}\sqrt{E_{J2}+E_{Jm}}}=\frac{C_m}{\sqrt{C_1+C_m}\sqrt{C_2+C_m}}.
\end{equation}
Note that Josephson energies can be tuned with relative precisions of $10^{-5}$ \cite{Wallraff13} or better in current experiments with transmons so that the tunneling terms in equation (\ref{rwatunnel}) can indeed be made negligible to the desired degree.
The rotating wave approximation applied for deriving equation (\ref{rwatunnel}) furthermore requires that,
\begin{equation}
E_{Jm}\ll\left(\frac{E_{C1}\overline{E}_{J1}^3\overline{E}_{J2}}{E_{C2}}\right)^{\frac{1}{4}}+\left(\frac{E_{C2}\overline{E}_{J2}^3\overline{E}_{J1}}{E_{C1}}\right)^{\frac{1}{4}}.
\end{equation}
The desired density-density interaction is contained in the nonlocal cosine interaction terms,
\begin{equation}
H_J=-\frac{E_{Jm}}{4}\phi_1^2\phi_2^2 \approx -2\sqrt{E_{C1}E_{C2}}\frac{E_{Jm}}{\sqrt{\overline{E}_{J1} \overline{E}_{J2}}}a_1^{\dag}a_1a_2^{\dag}a_2
\end{equation}
Here we have applied a rotating wave approximation and made use of the fact that occupations of the 2nd excited states of the qubits are vanishingly small which allows us to neglect terms of the form $a_1a_1a_2^{\dag}a_2^{\dag}+ \text{H.c.}$. Since the dynamics of our system is, for the initial state we consider, restricted to the subspace of at most one excitation per transmon we can write our Hamiltonian in terms of Pauli matrices,
\begin{equation} \label{supptwoqubitham}
H^{2\times2}= H_{sys} + H_g^{2\times2}
\end{equation}
where $H_{sys}$ is as in equation (1) of the main text with
$\omega_i = \sqrt{8E_{Ci}\overline{E}_{Ji}}/\hbar+\delta\omega_i$, where $\delta\omega_i$ accumulates all minor renormalizations of the transmon frequencies due to the nonlinear terms of the cosine potentials and
\begin{equation}
J= \frac{C_m}{2 \hbar \sqrt{C_1+C_m}\sqrt{C_2+C_m}} \sqrt{E_{C1}E_{C2}}.
\end{equation}
Since $C_{m}/(C_i+C_{m}) \ll 1$ we find the upper limit,
\begin{equation}
J < \frac{1}{10 \hbar} \, \sqrt{E_{C1}E_{C2}},
\end{equation}
for the strength of the qubit-qubit interaction.
In turn the couplings to the transmission lines read,
\begin{equation}
H_g^{2\times2}= 4 \I  \sum_{j=1,2} \frac{E_{Ci}}{\hbar}\left(\frac{\overline{E}_{Ji}}{2E_{Ci}}\right)^{\frac{1}{4}}n_{g,j}(\sigma^+_j-\sigma^-_j), 
\end{equation}
which become identical to equation (2) of the main text after applying a rotating wave approximation. 

A good performance of the transistor requires a high ratio $\tau_{2}/\tau_{1}$, see figure 2 of the main text. In order to nonetheless neglect the secondary coupling between transmission line 1 and transmon 2, we need to have $C_{m} \ll (C_{g2}/C_{g1}) C_{1} = \sqrt{\tau_{1}/\tau_{2}} \, C_{1}$ which also limits the achievable coupling $J$ but is compatible with experimental parameters up to very high values of $\tau_{2}/\tau_{1}$.

\end{document}